# Domain Adaptation-based Edge Computing for Cross-Conditions Fault Diagnosis

*Yanzhi Wang[a]; Jinhong Wu[a]; Chu Wang[a]; Qi Zhou[a]; Tingli Xie[a]\**

a. School of Aerospace Engineering, Huazhong University of Science and Technology, Wuhan 430074, China

*Abstract*—Fault diagnosis of mechanical equipment provides robust support for industrial production. It is worth noting that, the operation of mechanical equipment is accompanied by changes in factors such as speed and load, leading to significant differences in data distribution, which pose challenges for fault diagnosis. Additionally, in terms of application deployment, commonly used cloud-based fault diagnosis methods often encounter issues such as time delays and data security concerns, while common fault diagnosis methods cannot be directly applied to edge computing devices. Therefore, conducting fault diagnosis under cross-operating conditions based on edge computing holds significant research value. This paper proposes a domain-adaptation-based lightweight fault diagnosis framework tailored for edge computing scenarios. Incorporating the local maximum mean discrepancy (LMMD) into knowledge transfer aligns the feature distributions of different domains in a high-dimensional feature space, to discover a common feature space across domains. The acquired fault diagnosis expertise from the cloud-based deep neural network model (cloud model) is transferred to the lightweight edge-based model (edge model) using adaptation knowledge transfer methods. It aims to achieve accurate fault diagnosis under cross-working conditions while ensuring real-time diagnosis capabilities. We utilized the NVIDIA Jetson Xavier NX kit as the edge computing platform and conducted validation experiments on two devices. In terms of diagnostic performance, the proposed method significantly improved diagnostic accuracy, with average increases of 34.44% and 17.33% compared to existing methods, respectively. Regarding lightweight effectiveness, our method achieved an average inference speed increase of 80.47% compared to the comparative methods. Additionally, compared to the C-model, the parameter count of the E-model decreased by 96.37%, while the Flops decreased by 83.08%.

*Index Terms*—Deep learning, Edge computing, Fault diagnosis, Knowledge transfer, Lightweight neural network.

# 1. INTRODUCTION

Rotating machinery fault diagnosis involves detecting, analyzing, and assessing the fault status of mechanical equipment to identify and pinpoint underlying issues[1]. It plays a crucial role in ensuring production continuity, enhancing equipment reliability and operational efficiency, reducing production costs, extending equipment lifespan, and ensuring workplace safety, thereby providing robust support for industrial production[2, 3].

Over the years, deep learning techniques have emerged as powerful tools for fault diagnosis due to their ability to effectively analyze complex data patterns and make accurate predictions[4, 5]. Zhang et al. [6] utilized generative adversarial networks to learn the mapping between noise distribution and real mechanical vibration data, and then generated samples to balance and expand the dataset, addressing the problem of fault diagnosis under data imbalance. Zhao et al. [7] inserted soft thresholding as a non-linear transformation layer into the deep architecture to construct the Deep Residual Shrinkage Network. This model enhances the feature learning capability for high-noise vibration signals, achieving high fault diagnosis accuracy. Li et al. [8] proposed a two-stage transfer adversarial network, which can effectively separate multiple unlabeled new fault types from known fault types, and can be applied to detect various new faults in rotating machinery. It is worth noting that operating conditions are not constant, during the operational service of mechanical equipment. Its operation is accompanied by changes in factors such as rotation speed, load, etc., exhibiting characteristics of diversity[9]. This leads to significant differences in data distribution, indicating that the data are not from the same distribution. To address this issue, many studies have been conducted on fault diagnosis across different operating conditions [10, 11]. Lu et al. [12] employed a deep multi-scale CNN to extract multi-scale features from the measured raw vibration signals. Furthermore, they introduce a transformer block structure with a multi-head attention mechanism to establish connections between fault information and fault categories, demonstrating excellent fault feature spectrum recognition capability. Peng et al. [13] proposed a hybrid matching adversarial domain adaptation network, selecting multi-scale convolutional kernels. Simultaneously, embedding hybrid matching and confidence threshold in the adversarial network reduced conditional distribution differences. Ren et al. [14] proposed an unsupervised cross-task meta-learning strategy with a preference for distributional similarity, centered around a distribution distance weighting mechanism. The concept of maximum mean discrepancy was introduced simultaneously and used to measure the distribution distance. Chai et al. [15]extracted transferable and discriminative fault prototypes by simultaneously training a similarity-based discriminative module and a fault prototype adaptation module. Consequently, they learned transferable feature representations, reduced domain differences, and improved diagnostic performance on target conditions.

In addition to the aforementioned data-focused research, there has been a growing body of literature in recent years that addresses the deployment and implementation of methods[16, 17]. In particular, cloud-based approaches have been widely adopted in the field of industrial equipment health monitoring. In these methods, operational data is collected from the equipment, uploaded to the cloud, and inference results are obtained through models deployed on cloud servers[18]. However, due to the typically harsh operating environments of mechanical equipment, such methods face challenges such as inference latency and data security concerns, which significantly impact the effectiveness and safety of fault diagnosis[19].

To address some of the limitations associated with the centralized data processing model of cloud computing mentioned above. In recent years, with the widespread adoption of smart Internet-of-things devices and sensors, edge computing has gradually gained attention[20]. Edge computing pushes computational and data storage capabilities towards the network edge, allowing for data processing closer to where it is generated[21, 22]. This enables faster response times, more efficient data processing, and enhances data security. In terms of the theoretical foundation of edge computing, Shi et al. [23] systematically elucidated and researched the definition, characteristics, architecture, and principles of edge computing. Their proposed "cloud computing + edge computing" hybrid architecture offers a fresh perspective for addressing challenges such as big data processing, real-time computation, and distributed storage. Lu et al. [24] focused on edge computing techniques for machine fault diagnosis based on signal processing. Specifically, they investigate the utilization of lightweight algorithms and specialized hardware platforms tailored for application in fault diagnosis procedures. Wang et al. [25] present an effective asynchronous federated learning approach that addresses resource constraints at the edge, reduces communication overhead, and accelerates training in heterogeneous edge environments. Janjua et al. [26]present an abnormal monitoring system that can apply unsupervised machine learning techniques directly on gateways positioned at the edge. Utilizing this edge computing framework for data analysis and fault diagnosis near the target devices enables instantaneous feedback on outcomes. In the context of real-time fault diagnosis, edge computing facilitates the processing of data closer to the data source, thus mitigating the latency associated with data transmission and enabling swift fault diagnosis[27]. Simultaneously, deep learning techniques leverage data on edge devices for real-time pattern recognition and anomaly detection, facilitating rapid fault diagnosis and localization[28]. The

integration of edge computing and deep learning yields several advantages[29, 30]:

**(1) Real-time capability:** Edge computing enables fault diagnosis at the point of data generation, thereby achieving real-time processing.

**(2) Precision capability:** Leveraging deep learning within the edge computing framework allows for real-time analysis of data at the source, avoiding misdiagnosis and missed detections caused by deteriorated data quality during transmission.

**(3) Data Security:** Edge computing enables the processing of data locally, without the need to transmit it to centralized servers or cloud platforms. This reduces the exposure of sensitive information to potential security threats that may exist in transit or at remote data centers.

The fusion of edge computing and deep learning offers more efficient and precise solutions for real-time monitoring and diagnosis tasks. However, current research on edge computing mainly focuses on method principles and application frameworks[24]. There is less in-depth research on specific problems encountered during application. For fault diagnosis methods under cross-operating conditions, typically require significant computational resources for both training and inference, which often exceeds the capabilities of many edge devices. Therefore, based on the principles of deep learning and edge computing, conducting research on fault diagnosis methods under cross-operating conditions is of great significance. This paper proposes a lightweight intelligent fault diagnosis framework based on domain adaptation tailored for edge computing scenarios. By introducing domain adaptation learning into fault diagnosis, different domain feature distributions are aligned in a high-dimensional feature space, enabling the discovery of a common feature space for knowledge transfer between different domains. The main contributions of this paper are as follows:

**(1)** The fault diagnosis process is constructed around the principles of edge computing. Optimal weights are trained and allocated based on samples on the cloud server to ensure high diagnostic accuracy of the cloud model (C-model). Subsequently, the lightweight edge models (E-model) are utilized at the edge computing end to promptly diagnose actual operational conditions.

**(2)** The proposed knowledge transfer method is employed to transfer the acquired fault diagnosis expertise from the C-model to the E-model. It aligns the feature distributions of different domains in a high-dimensional feature space, enabling the transfer of fault diagnosis knowledge learned under a single operating condition to actual operating conditions.

**(3)** Based on domain adaptation, it effectively guides the E-model to better comprehend and diagnose complex real-world knowledge, thereby achieving outstanding fault diagnosis results.

 **(4)** Comparative tests and annealing experiments were conducted on two experimental platforms. The results indicate that the proposed method can effectively perform domain adaptation learning across different operating conditions. The E-model exhibits better diagnostic performance and better lightweight metrics.

The paper follows this structure: Section 1 introduces the background of the proposed method. Section 2introduces the technical basis of the proposed method. Section 3 elaborates on the proposed method. Section 4 illustrates experiments and evaluations. Lastly, Section 4 outlines the conclusions drawn from the study and suggests areas for future research.

## 2. PROPOSED METHOD

### 2.1. Workflow of the Proposed Method

This study presents an intelligent fault diagnosis method based on domain adaptation to address the issue of cross-condition fault diagnosis in rotating machinery in edge computing scenarios. The proposed method comprises three stages: cloud-based model training, knowledge transfer at the edge, and real-time fault diagnosis at the edge. Fig. 1 illustrates the main workflow of this method.

In the cloud, firstly, numerous normal samples and samples of each fault category under a single working condition are obtained through simulation experiments. Then, the C-model is trained based on the above large number of samples. After iterative training obtain the optimal weights of the model, which are sent down to the edge end. At the edge, the excellent feature extraction capability of the C-model is transferred to the lightweight E-model through knowledge transfer, using a small number of samples. Additionally, the diagnostic capability from the source domain (Condition 0) is generalized to the target domain (Condition N). Then, the E-model is deployed and run. The method achieves fast and accurate fault diagnosis near the device by inputting real-time operational data.

The key aspect of the proposed method is in the domain adaptation technique, which enables the transfer of knowledge from a complex C-model to a simple E-model and the adaptation from the single condition to the various operating conditions.

### 2.2. Model Structure

The proposed method involves two network architectures: the C-model, which is trained in the cloud and provides fault knowledge at the edge, and the E-model, which is deployed at the edge for real-time diagnosis. In the operational framework of the proposed method, the C-model is expected to have a deep network architecture to ensure its ability to effectively capture and represent data features, thereby achieving accurate prediction capabilities. Meanwhile, the E-model should have a smaller model size and lower computational complexity while maintaining sufficient feature learning capabilities. In addition, it should exhibit strong generalization capabilities to enable knowledge transfer from the C-model and facilitate fast and effective inference under various operating conditions on edge devices.

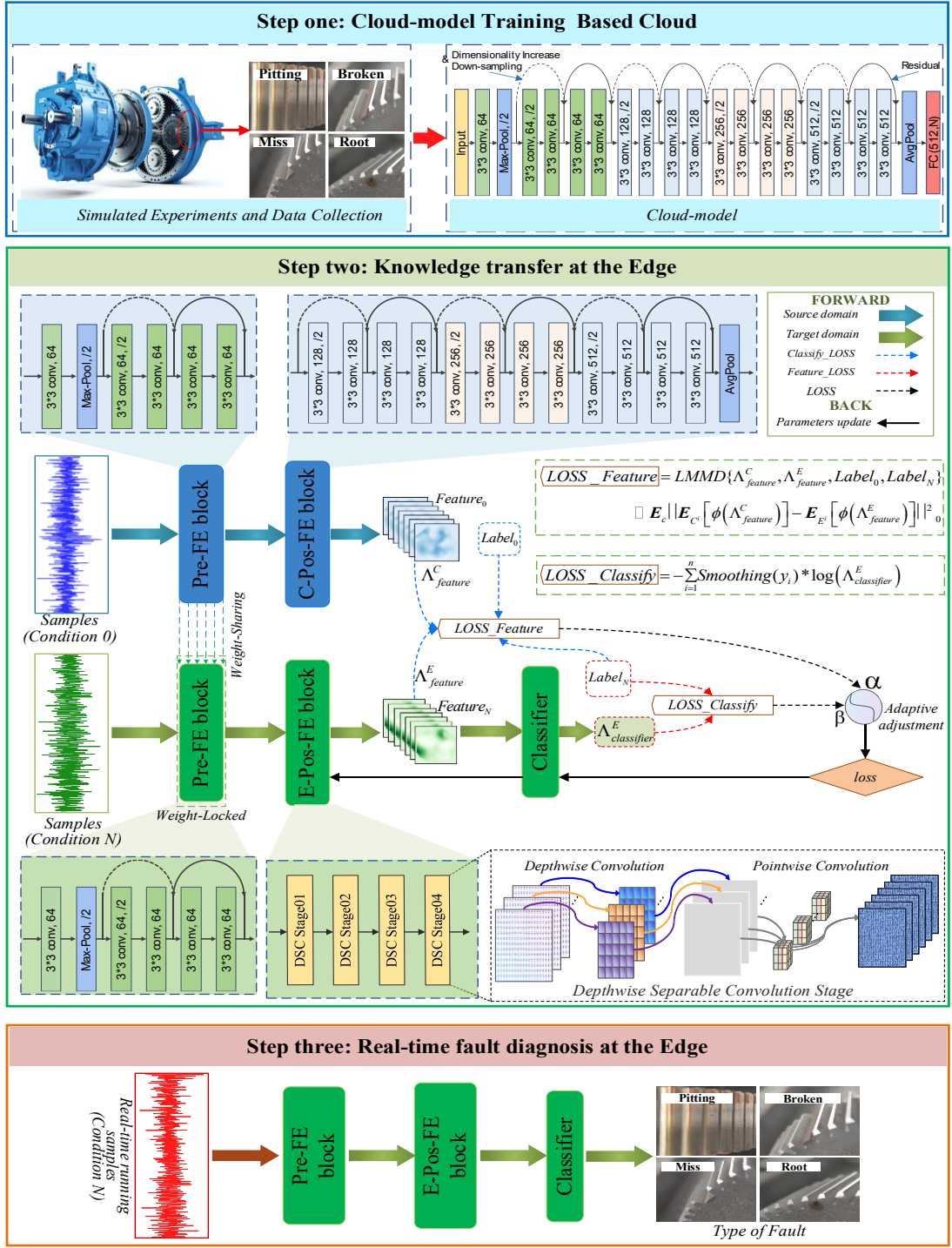

**Fig. 1 The specific workflow of the proposed method**

Fig. 1 illustrates the composition of the C-model, which mainly consists of a Pre-Feature Extractor block (Pre-FE block), a Posterior Feature Extractor block (C-Pos-FE block), and a classifier. The feature extractor extensively utilizes residual connection modules. Residual modules can be implemented using skip connections, where the input to a unit is added directly to the output of the unit, followed by activation. This approach not only ensures the depth of the model and improves the feature learning performance of the model, but also mitigates the problem of gradient disappearance, and reduces the training difficulty.

The structure of the E-model is similar to that of the C-model, consisting of a Pre-Feature Extractor block (Pre-FE block), a Posterior Feature Extractor block (E-Pos-FE block), and a classifier. The structure of the Pre-FE block is identical to that of the C-model. This design allows weight sharing during knowledge transfer at the edge so that the E-model has some of the feature extraction capabilities of the C-model and can also reduce the amount of calculation. The E-Pos-FE block uses four depth-wise separable convolution stages, as shown in Fig. 1. This approach using convolution ensures that features are extracted while reducing the number of parameters and enhancing computational efficiency at the same time.

## 2.3. Knowledge Transfer Methodology

The C-model is initially iteratively trained on the cloud computing side. This enables the C-model to effectively capture and represent features of the samples, acquiring accurate recognition capabilities for single operating conditions. Subsequently, there is a need to transfer the diagnostic capability from the large and complex neural network (C-model) to a smaller and simpler neural network (E-model), while also generalizing the diagnostic capability from single operating conditions to other conditions. This paper proposes a novel knowledge transfer method aimed at optimizing model size, enhancing computational efficiency, and aligning the diagnostic capability of the E-model with real-world operating conditions. The proposed knowledge transfer method, which consists of two stages, is illustrated in Fig. 1.

When constructing both the C-model and the E-model, the same Pre-FE block was employed as the front feature extractor. During knowledge transfer, weight sharing was conducted on the Pre-FE block first. By employing a weight-sharing mechanism, the E-model can fully utilize the feature extraction capabilities of the C-model. Specifically, this implies that the E-model effectively inherits and shares the feature representations learned by the C-model. This process reduces the computational cost of training this part of the feature extraction capability on edge devices, effectively alleviating the computational burden.

After completing the above weight-sharing operation, lock the weight of the Pre-FE block in the E model. Next, we train the E-Pos-FE block and classifier via Algorithm 1. Specifically, samples from Condition 0 are first input into the feature extractor of the C-model, resulting in feature representation $\Lambda_{classifier}^{C}$. While samples from Condition N are input into the feature extractor of the E-model, resulting in feature representation $\Lambda_{classifier}^{E}$. Next, based on $\Lambda_{classifier}^{C}$, $\Lambda_{classifier}^{E}$, and the corresponding labels of the samples, the differences in the distribution of related subdomains are calculated using LMMD and denoted as $LOSS\_Feature$. It is then incorporated as one of the components in the model optimization objective.

Furthermore, the classifier of the E-model outputs the predicted categories of samples from Condition N, which undergo label smoothing to obtain $\Lambda_{classifier}^{E}$. Combined with the true labels $Label_N$, $LOSS\_Classify$ is calculated using the cross-entropy loss function, serving as another component of the model optimization objective. Assign weights α and β to $LOSS\_Feature$ and $LOSS\_Classify$, respectively. Calculate loss using the formula: $loss = \alpha * LOSS\_Feature + \beta * LOSS\_Classify$, and utilize it as the final objective for minimization optimization. By employing backpropagation, optimize the parameters of the neural networks in the E-Pos-FE block and the Classifier to obtain the optimal weights of the E-model.

The transfer of fault diagnosis knowledge from the cloud computing model to the E model is achieved. Additionally, the optimal weights for adapting the edge model to real-world operating conditions are obtained using a small number of actual condition samples.

**Algorithm 1**: Training process of E-Pos-FE block and classifier.

| | |
|---|---|
| Input: | Raw signal samples for different fault categories with normal and fault states. |
| Output: | The optimal model parameters of E-Pos-FE block and classifier: $\hat{\theta}$. |
| Initialize: | Fine-tuning-data set: $s^0, l^0$ (sample and label of Condition 0), $s^N, l^N$ (sample and label of Condition N). |
| 1. | for $i = 1:iterations$ |
| | Obtain the feature representation of the C-model and E-model |
| 2. | $\Lambda_{feature}^{C} = [Pre\text{-}FE \rightarrow C\text{-}Pos\text{-}FE](s^0)$, |
| | $\Lambda_{feature}^{E} = [Pre\text{-}FE \rightarrow E\text{-}Pos\text{-}FE](s^N)$ ; |
| 3. | Calculate one of the optimization objectives: $LOSS\_Feature = LMMD\{\Lambda_{feature}^{C}, \Lambda_{feature}^{E}, l^0, l^N\}$ ; |

4. Get the smoothed predicted label: $\Lambda^E_{classifier} = Smoothing\{Classifier^E(\Lambda^E_{feature})\}$;
5. Calculate another optimization objective:
$$LOSS\_Feature = -\sum_{i=1}^{n} Smoothing(Label_N) * \log\left(\Lambda^E_{classifier}\right) ;$$
6. Calculate the *loss*: $loss = \alpha * LOSS\_Feature + \beta * LOSS\_Classify$;
7. $update(\hat{\theta})$
8. end for

## 2.4. The Approach of Domain Adaptive

In the knowledge transfer process mentioned above, parameters α and β are set to adjust the weights of *LOSS_Feature* and *LOSS_Classify*. Furthermore, adjusting the influence of the guidance role of the E-model's diagnostic effectiveness under actual operating conditions and the feature recognition capability of the C-model on the loss during the training process. However, in practical operations, it is challenging to quantify the differences between various conditions manually. Therefore, the proposed method adopts domain adaptation to adjust parameters α and β, thereby achieving better adaptive diagnosis across operating conditions.

The approach of domain adaptive primarily employs two key elements: the computation of *LOSS_Classify* using LMMD, and the adaptive adjustment of the weights for *LOSS_Feature* and *LOSS_Classify* to obtain the total optimization objective *loss* during the knowledge transfer process.

The optimization objective *LOSS_Classify* aims to minimize the distribution differences between the target condition and the source domain condition. Enabling the proposed method to extract fault features with high correlation across operating conditions. The LMMD method is utilized to compute *LOSS_Classify* during the training process.

The total optimization objective *loss* in the knowledge transfer process is obtained by adaptively adjusting the weights of *LOSS_Feature* and *LOSS_Classify*. First, calculate the gradient of *LOSS_Feature* and *LOSS_Classify* on the output value $\Lambda^E_{classifier}$ of the E-model respectively:

$$Grad\_Classify = Gradient(LOSS\_Classify, \Lambda^E_{classifier}) \quad (1)$$

$$Grad\_Feature = Gradient(LOSS\_Feature, \Lambda^E_{classifier}) \quad (2)$$

Then, the weights of *LOSS_Feature* and *LOSS_Classify* are calculated separately

for the dynamic calculation of *loss:*

$$\alpha = \frac{w_a}{w_a + w_b + \eth} \cdot \frac{l_a + l_b}{l_a + \eth} \quad (3)$$

$$\beta = \frac{w_b}{w_a + w_b + \eth} \cdot \frac{l_a + l_b}{l_b + \eth} \quad (4)$$

where $l_a$ and $l_b$ are the L2 paradigms of *LOSS_Feature* and *LOSS_Classify*, $w_a$ and $w_b$ are the L2 paradigms of *Grad_Feature* and *Grad_Classify*, respectively, and $\eth$ is a very small value for avoiding division by zero error. Further, the total optimization objective *loss* during knowledge migration of the proposed method is calculated:

$$loss = \begin{cases} \alpha * LOSS\_Feature + \beta * LOSS\_Classify, & epoch \leq 0.9 * num\_epoch \\ LOSS\_Classify, & epoch > 0.9 * num\_epoch \end{cases} \quad (5)$$

Therefore, the proposed method adjusts parameters α and β, adaptively. Through the backpropagation optimization process of loss affecting the edge model, the adaptation to actual working conditions is achieved.

## 3. CASE STUDIES

### 3.1. Experimental background

The research utilized both a computer and a Jetson Xavier NX kit to conduct case studies. The computer served as the cloud device, equipped with an Intel Core i7-11700K@3.60GHz processor, an NVIDIA GeForce RTX 3070 GPU, and 32.0GB of RAM. Meanwhile, the Jetson Xavier NX kit is used as the edge device and features an NVIDIA Carmel ARM CPU, an NVIDIA Volta GPU with 48 tensor cores, and 8.0 GB of RAM. Operating on the Windows platform, the cloud device hosted the dependency environment essential for the models. Conversely, the edge device utilized Docker, an open-source application container engine, for deployment purposes. The models were developed using Python 3.9 and PyTorch 1.11.0.

Under the aforementioned experimental conditions, we conducted case studies for the proposed method using two datasets. We compared the proposed method with conventional approaches, using Rotating Machinery Fault Simulation (RMFS), This comparison aimed to substantiate the fault diagnosis efficiency and lightweight performance in edge computing scenarios. Furthermore, to validate the efficacy of the proposed domain-adaptive knowledge transfer method, a series of ablation experiments were executed using the Drivetrain Dynamics Simulator (DDS).

## 3.2. CASE 01
### 3.2.1. Experimental Method

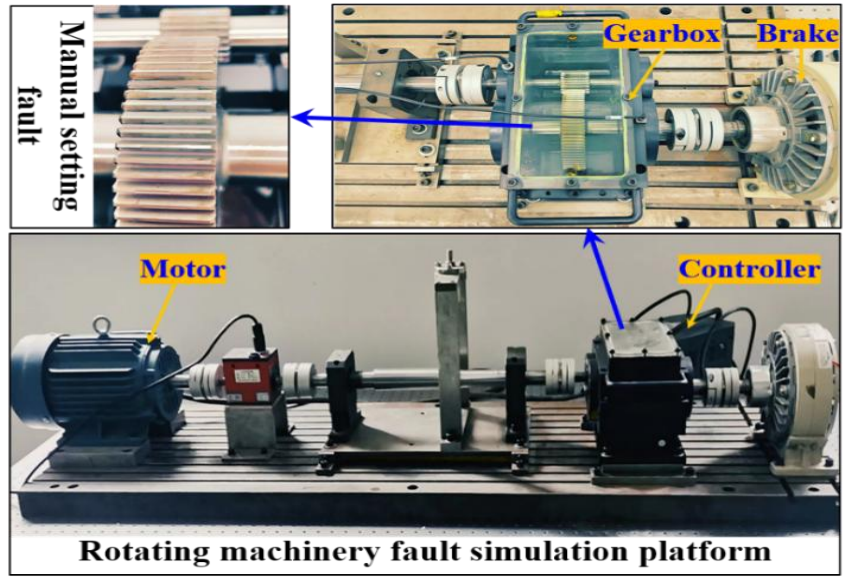

**Fig. 2** Experimental equipment of RMFS.

**Table 1.** Fault setting method of RMFS.

| Fault label | Fault type | Fault processing method |
|---|---|---|
| 0 | Normal | - |
| 1 | Broken | Cut off 1/2 of a single whole tooth |
| 2 | Miss | A tooth completely missing |
| 3 | Root | A crack at the gear root<br>Crack specification: width 0.2 mm, depth 1mm |
| 4 | Pitting | Make four pitting corrosion by an electric spark<br>Specification of single place: five times of electric spark |

Case 01 was performed using the customized RMFS platform developed by Huazhong University of Science and Technology, as shown in **Fig. 2**. The simulation platform includes an electric motor, controller, bearings, gearbox, and brake[33]. It enables the introduction of various faults in the bearings and gearbox, and the operational conditions can be adjusted by modifying input power and load. The ZDY80 parallel-axis gearbox was chosen as the subject of investigation in this study. The gearbox can simulate five different fault types, as detailed in **Table 1**. Experiments were conducted with input powers of 1800W and 3000W, and loads set at 0, 50 n•m, and 100 n•m.

**Table 2.** The experimental groups of CASE 01.

| Source domain \ Target domain | Load:50 n•m; Power:1.8kw | Load:50 n•m; Power:3.0kw | Load:100 n•m; Power:1.8kw | Load:100 n•m; Power:3.0kw |
|---|---|---|---|---|
| Load:0; Power:1.8kw | A1 | A2 | A3 | A4 |
| Load:0; Power:3.0kw | B1 | B2 | B3 | B4 |

Accelerometers were placed on the vertical axis of both the high-speed and low-speed termini shafts of the parallel gearbox to capture acceleration signals along the x, y, and z axes at these respective locations. A sampling method was implemented using a sliding window with a consistent window length of 1024 without window overlap.

In the experiment, data under the zero-load condition was stored on cloud servers, while data under two different load conditions was collected based on edge devices. The experiments are based on two cloud-based sample sets and four edge-based sample sets. To fully verify the benefits of the proposed method, eight sets of experiments are conducted, as shown in Table 2. In the cloud-based samples, 250 samples were randomly selected for each fault category as the training dataset ($D_{training}$).

For the edge-based samples, 10 samples were randomly extracted from each fault category based on edge device-based data, and merge the 10 samples taken in the $D_{training}$ to form the fine-tuning dataset ($D_{fine-tuning}$), and an additional 100 samples were taken as the test dataset ($D_{test}$). To validate the diagnostic performance of the proposed method and its applicability in edge computing scenarios, the following models were selected for comparative experiments:

**(1) Mobilenet:** MobileNet is tailored for deployment on mobile and embedded devices. It uses depth-wise separable convolutions to minimize computational overhead while maintaining high accuracy[31].

**(2) ShuffleNetV2:** ShuffleNetV2 builds upon the foundation of the classic lightweight network ShuffleNetV1 block, incorporating pointwise group convolutions and bottleneck-like structures. The introduction of channel shuffle operations enables information exchange between channels from different groups, thereby enhancing the final accuracy[34].

**(3) Xception:** Xception is an important architecture in deep learning. The model demonstrates superior performance with a relatively smaller model size, making it highly valuable for deep learning applications in resource-constrained environments[35].

**(4) C-model-transfer:** To adapt to the scenarios of variable working conditions, transfer learning is used to train the C-model. First, the C-model is trained based on the $D_{training}$. Then, the feature extraction layer of the C-model is locked, and the classifier of the C-model is trained using the $D_{fine-tuning}$. Finally, testing is performed on the $D_{test}$.

**(5) E-model-transfer:** The method of transfer learning is used to train the E-model. The specific steps are the same as the **C-model-transfer** method.

**(6) Proposed method:** Utilizing $D_{training}$, the C-model is trained on cloud servers, and the optimal weights of the C-model are downloaded to edge devices. Using $D_{fine-tuning}$, adaptive knowledge transfer is conducted to derive the weights of the E-model. Subsequently, the E-model is tested on $D_{test}$.

MobileNet and ShuffleNetV2 are efficient lightweight network models recently. The proposed method is compared with these two networks to evaluate its lightweight efficiency. Xception has been a high-efficiency convolutional neural network in recent years. The proposed method is compared with the method to evaluate their diagnostic capabilities. MobileNet, ShuffleNetV2, and Xception are trained on the data sets composed of $D_{training}$ and $D_{fine-tuning}$, and tested on $D_{test}$. The proposed method is also compared with E-model-transfer and C-model-transfer to measure the knowledge transfer efficiency of the proposed method.

### 3.2.2. Results & Discussion

In the experiment, the samples input to each network are reshaped into a matrix of size [6, 56, 56], respectively. During training the bath size is set to 32. CosineAnnealingLR is used to modify the learning rate (LR) of the network during training, the method adjusts the LR by the cosine function, which can make the learning take the lead in a slow decline, then accelerate the decline, and then slowly decline again. This method yields advantageous results in accelerating model convergence and enhancing model efficacy.

To comprehensively compare our proposed method with other classical approaches, we conducted 5 repeated experiments. The test accuracies of each model in different groups were recorded, as shown in Fig. 3, where error bars represent the standard deviation of the results. Furthermore, the average diagnostic accuracy of each model was calculated and summarized in Table 3 for quantitative analysis. It is evident that our proposed method outperforms the comparative methods, indicating its strong diagnostic capability under varying conditions. Specifically, apart from the average diagnostic accuracy of 90.08% in Group B3, the method consistently achieved accuracies above 95% in the remaining 7 groups. Moreover, the diagnostic results showed relative stability compared to the comparative methods. Regarding MobileNet, except for the result of 78.91% in Group A4, the accuracies in the other groups were above 80% but still below 90%. This phenomenon suggests the outstanding feature extraction capability of depth-wise separable convolutions. In addition, MobileNet showed less overfitting to the source domain during the experiment due to its lightweight design with fewer parameters. Consequently, it demonstrated good performance in cross-condition testing. ShuffleNetV2 yielded results ranging from 60% to 80%, with poor stability, showcasing diagnostic capabilities inferior to MobileNet. Xception, as a classic deep neural network with a larger parameter count, exhibited poor adaptability during cross-condition fault diagnosis experiments. C-model-transfer and E-model-transfer, as classic transfer learning methods, both showed test results below 50% in each group, with poor stability. This suggests that despite the availability of a small amount of labeled target domain data in the experiment, the limited sample size constrained the knowledge gained, leading to suboptimal performance.

Table 3. The average test accuracy of different models in Case One (%).

| Model | A1 | A2 | A3 | A4 | B1 | B2 | B3 | B4 |
|---|---|---|---|---|---|---|---|---|
| Mobilenet | 88.43 | 86.89 | 86.37 | 78.91 | 88.79 | 85.93 | 87.06 | 85.01 |
| ShuffleNetV2 | 61.98 | 67.34 | 60.89 | 62.10 | 79.60 | 67.02 | 63.31 | 71.45 |
| Xception | 69.10 | 77.02 | 66.05 | 79.04 | 76.25 | 71.77 | 71.21 | 81.66 |
| C-model-transfer | 45.12 | 40.08 | 33.564 | 46.33 | 53.75 | 37.91 | 41.69 | 45.53 |
| E-model-transfer | 42.14 | 40.40 | 37.54 | 43.79 | 38.67 | 46.09 | 28.48 | 42.42 |
| **Proposed method** | **98.37** | **95.12** | **97.24** | **98.10** | **95.36** | **97.58** | **90.08** | **99.02** |

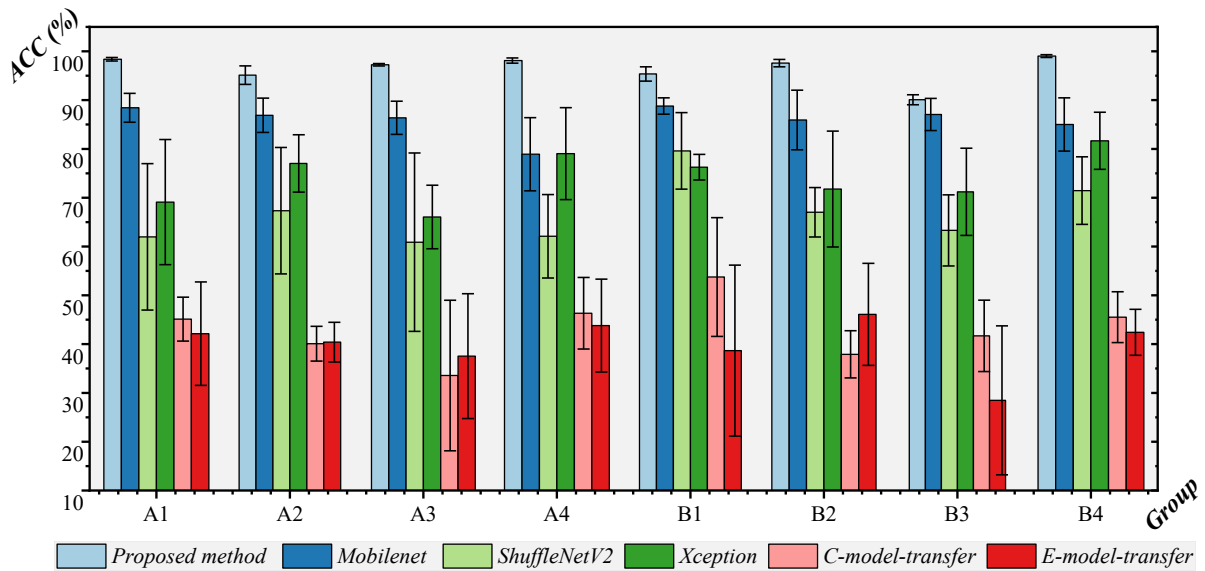

Fig. 3 The test accuracy of different models in Case One.

In addition to assessing the models' performance in classification accuracy and feature extraction capabilities for the specific task, an imperative consideration lies in the evaluation of their network complexity. The parameters associated with the models hold significance, particularly in the context of edge computing scenarios. Specifically, the following parameters of different models are compared in this section:

**(1) Params:** Denoting the total number of trainable parameters during model training, this metric serves as a gauge for the model's size and contributes to the calculation of space complexity.

**(2) Memory:** This parameter signifies the volume of memory required for model inference, offering insights into the model's memory utilization.

**(3) Floating Point Operations (FLOPs):** Expressing the theoretical volume of floating-point arithmetic operations within the neural network, FLOPs offer insight into the computational workload imposed by the model.

**(4) Inference Time:** Referring to the duration taken for the inference of a single sample on the edge-end device, this parameter is crucial for assessing real-time processing capabilities.

Table 4 showcases the values of Params, Memory, FLOPs parameters, and the average inference time per single sample for each model. The statistical calculations for Params, Memory, and FLOPs parameters were executed using torchstat, a lightweight neural network analyzer based on PyTorch. Deploy the model on the edge device, perform 1000 inferences take the average, record it as one result, repeat 10 times, and take the average to get the final inference time. It is noteworthy that the GPU underwent equal-intensity warm-up procedures before each speed measurement. Fig. 4 visually depicts the results of this comprehensive evaluation.

**Table 4. Test results of the complexity of each model.**

| Methods | Params | Memory (MB) | Flops (MFlops) | Inference time (ms) |
|---|---|---|---|---|
| Mobilenet | 610,213 | 1.05 | 7.31 | 4.75 |
| ShuffleNetV2 | 5,495,052 | 1.06 | 10.4 | 38.29 |
| Xception | 20,818,061 | 2.60 | 88.94 | 23.02 |
| C-model-transfer | 11,177,349 | 1.47 | 89.17 | 16.48 |
| Proposed method | **405,957** | **1.20** | **15.09** | **4.03** |

The parameter amount of the proposed method is of the same order of magnitude as that of MobileNet. Notably, the parameter count of the proposed method is minimal, approximately

7.38% of another lightweight model, ShuffleNetV2. The reduction in parameter count enhances the efficiency of the model training process. Specifically, a relatively smaller model size contributes to shorter training times, facilitates the identification of suitable parameter values, and improves training tability. Additionally, in cross-condition scenarios, the model's lower parameter count indicates a comparatively weaker fitting ability to train data, thereby assisting in preventing overfitting. The Memory and FLOPs of the proposed method align with those of MobileNet and ShuffleNetV2. Notably, the proposed method's Memory and FLOPs are 46.15% and 16.97% of Xception, respectively. The lower Memory and FLOPs signify reduced hardware memory usage and lower computational resource requirements, making the model more amenable to embedding in resource-constrained edge devices. Regarding inference speed, the proposed method achieves optimal performance. Specifically, the inference time of the proposed method is 10.52% of ShuffleNetV2 and 17.52% of Xception, demonstrating the ability to rapidly obtain diagnostic results at the edge.

It is noteworthy that the proposed method's parameter count, Memory, FLOPs, and inference time are 3.63%, 81.63%, 16.92%, and 24.45% of C-model-transfer, respectively. This suggests that, despite both undergoing training on a cloud server for the C-model, training the E-model using the proposed method results in a more lightweight model compared to classical transfer learning approaches. Consequently, this achieves reduced hardware overhead and faster inference speeds.

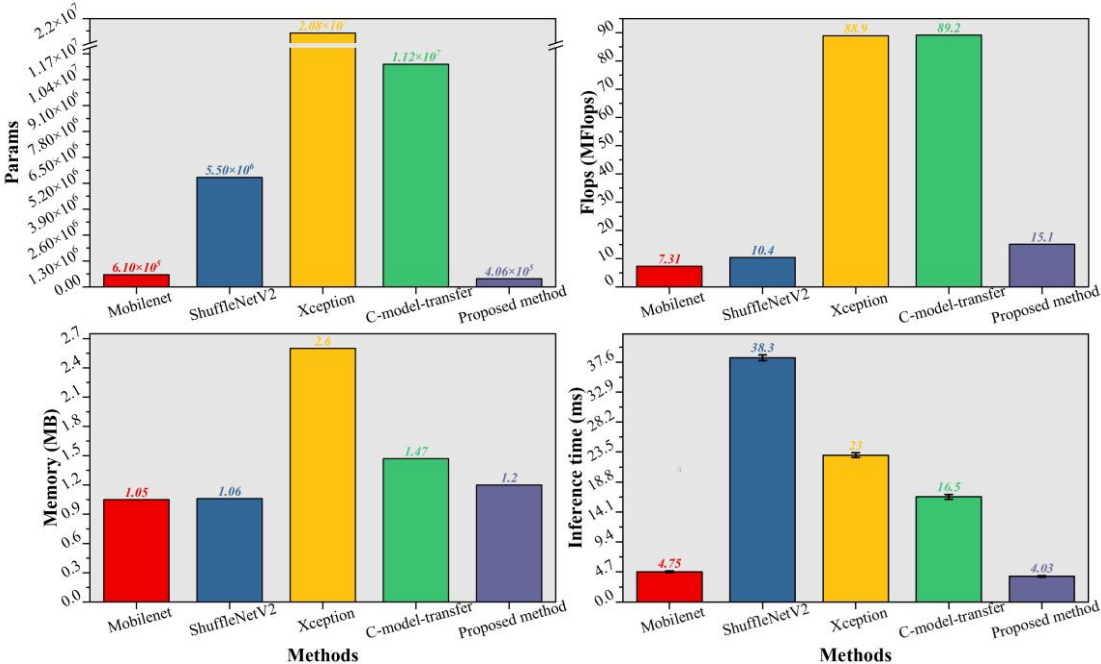

**Fig. 4 Visualization of complexity metrics for each model**

## 3.3. Case 02
### 3.3.1. Experimental Method

This parameter signifies the volume of memory required for model inference, offering insights into the model's memory utilization.

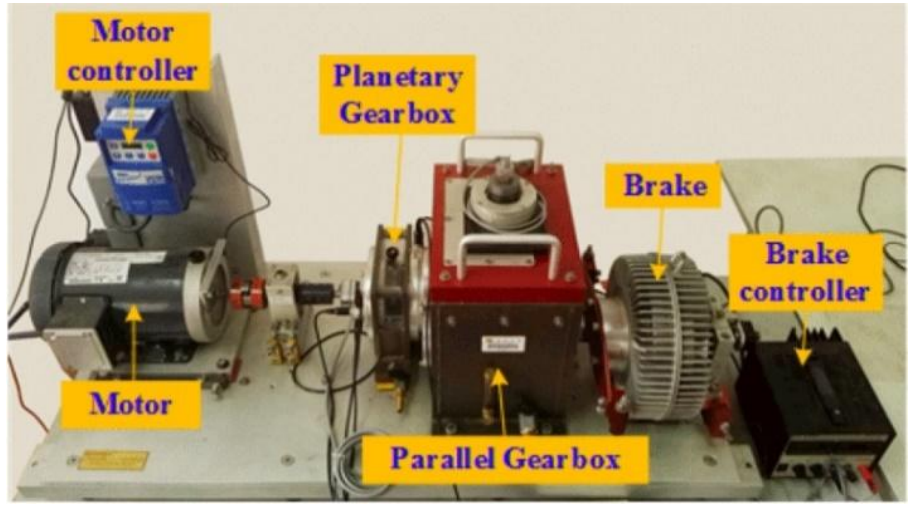

**Fig. 5** Experimental equipment of DDS.

The experimental data for Case 02 was obtained using the Dynamic Drive System (DDS) at Southeast University, China[36]. The simulation platform illustrated in **Fig. 5**. enables the emulation of both healthy and various faulty operating conditions in the bearing and gearbox, including Health, Gear teeth broken, Gear teeth missing, Root crack, Surface pitting corrosion, Bearing ball, Bearing comb, Bearing inner ring, and Bearing outer ring. These conditions are denoted as 0-8 during the experiments. The experimental device consists of two working conditions: 1200 rpm with 0Nm load and 800 rpm with 7.32Nm load. Two groups of experiments are conducted to fully test the effectiveness of the method:

**C:** The data for the working condition of 1800rpm with 7.32Nm load is saved in the cloud server for model training, and the data for the working condition of 1200rpm with 0Nm load is used for testing in edge device;

**D:** The data with 0Nm load with 1200rpm is stored in the cloud server for model training and the data with 7.32Nm load with 1800rpm is used for testing in edge device.

Acceleration signals along the x, y, and z axes were acquired by installing acceleration sensors on the parallel and planetary gearboxes of the experimental platform. Subsequently, a sliding window approach was employed for sampling, with a fixed window length of 1024 and a zero-window overlap. In the cloud server-based data, 1023 samples are randomly selected for each fault category as the training sample set $S_{training}$. Randomly select 20 samples from each fault category based on edge device-based data, and merge the 20 samples taken in the $S_{training}$ to form a fine-tuning sample set ($S_{fine-tuning}$). In addition, 100 samples in each fault category based on edge device data are taken as a test sample set ($S_{test}$). In the ablation experiments in this section, the following reduced versions of the models are selected for comparison:

(1) **W/O-Domain-adaptation:** On the cloud server, the C-model is trained based on $S_{training}$. After completing the weight sharing of the Pre-FE block at the edge, the remaining part of the E-model is trained based on $S_{fine-tuning}$. Specifically, the cross-entropy loss function is used to calculate the loss on the edge device, and the weight of the E-model is updated through backpropagation. This method can measure the effect of the domain adaptation method adopted in this paper.

(2) **W/O-Adaptation-adjustment:** Train the C-model based on $S_{training}$ on the cloud server. After sharing the weight of the Pre-FE block at the edge, Based on $S_{fine-tuning}$, the knowledge migration method proposed in this article is used to migrate the knowledge of the C-model to the E-model. In particular, *LOSS_Feature* and *LOSS_Classify* are numerically added to calculate LOSS, rather than adaptive adjustment. This method can measure the effect of the adaptive adjustment method adopted in this article.

(3) **Proposed method:** Train the C-model on the cloud server based on $S_{training}$. At the edge, based on $S_{fine-tuning}$, the adaptive knowledge transfer method proposed in this article is used to migrate the knowledge of the C-model to the E-model. The prediction effect of the model is tested based on $S_{test}$.

### 3.3.2. Results & Discussion

In the experiment, the input samples for each model are reshaped into a matrix of size [6, 32, 32]. In training, the batch size is set to 32, considering the convergence of the model and memory usage. Cosine Annealing LR is used to adjust the LR during training. To ensure the robustness and reliability of the results, C and D groups of experiments were independently repeated 5 times and the average values were used for subsequent analysis.

To improve the persuasiveness of the ablation experiments, we conducted five repetitions of each experiment. We recorded the test accuracies of each model in the two groups, as shown in Fig. 6. The error bars represent the standard deviation of the results, and the table below displays the average diagnostic accuracy of the methods. The results in the figure exhibit a stair-step distribution, and our proposed method significantly outperforms the two comparison methods. Specifically, in both sets of experiments, the proposed method shows an average improvement of 19.76% and 11.02% over the W/O-Domain-adaptation and W/O-Adaptation-adjustment methods, respectively. These results indicate that the proposed method, which utilizes self-adaptive adjustment of LOSS weights and pre-adaptation methods, effectively enhances the fault diagnostic capability under variable operating conditions. Meanwhile, the stability of the two comparison methods is relatively poor. This is particularly evident in the W/O-Domain-adaptation method, where the accuracy deviation in the C group experiments exceeds 30%. This suggests that due to the limited number of target domain samples provided in the experiment, the comparison methods struggle to acquire sufficient diagnostic knowledge, leading to unstable results.

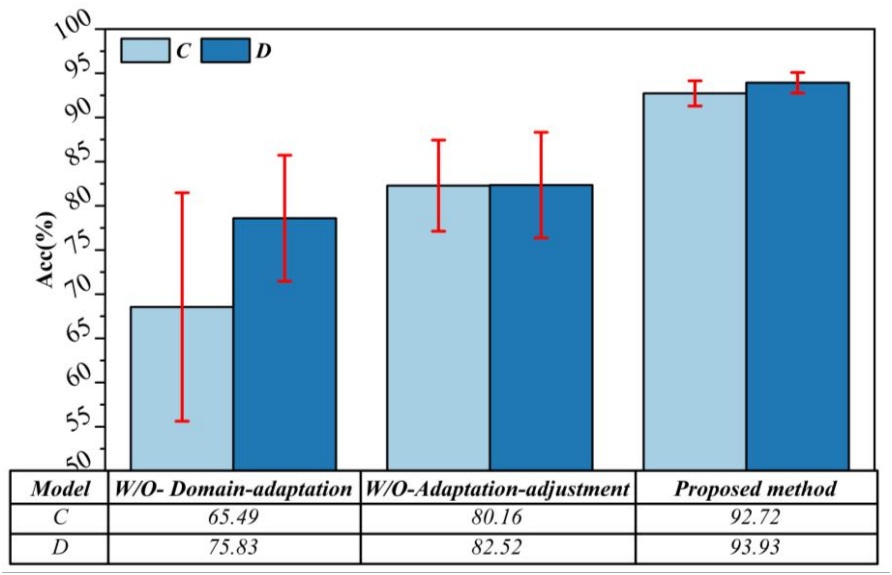

Fig. 6 The test accuracy of different models in Case Two.

To conduct an in-depth analysis of the specific classification predictions of different models under varying operating conditions, we chose to generate a confusion matrix for the test results of Group C, as illustrated in Fig. 7. The matrix displays the median corresponding diagnostic results for each model across five repeated experiments. The vertical axis represents the actual labels of the samples, while the horizontal axis represents the predicted results. In the diagnostic results of W/O-Adaptation-adjustment, three groups achieved an accuracy exceeding 95%, while four groups had diagnostic results below 80%. Specifically, the diagnoses for 'Miss' and 'Inner' were below 60%, significantly impacting the accurate diagnosis of these two categories. In the diagnostic results of W/O-Domain-adaptation, the diagnostic accuracy for all categories was below 80%, with diagnoses for 'Miss', 'Root', and 'Inner' falling below 50%. This method exhibited clear deficiencies in diagnostic capability, making it challenging to be effectively applied in practical engineering scenarios. In contrast, our proposed method demonstrated superior performance in the Group C test results, with six groups achieving accuracy rates exceeding 95%. This indicates that our proposed method can acquire sufficient knowledge in experiments, enabling more precise diagnostics.

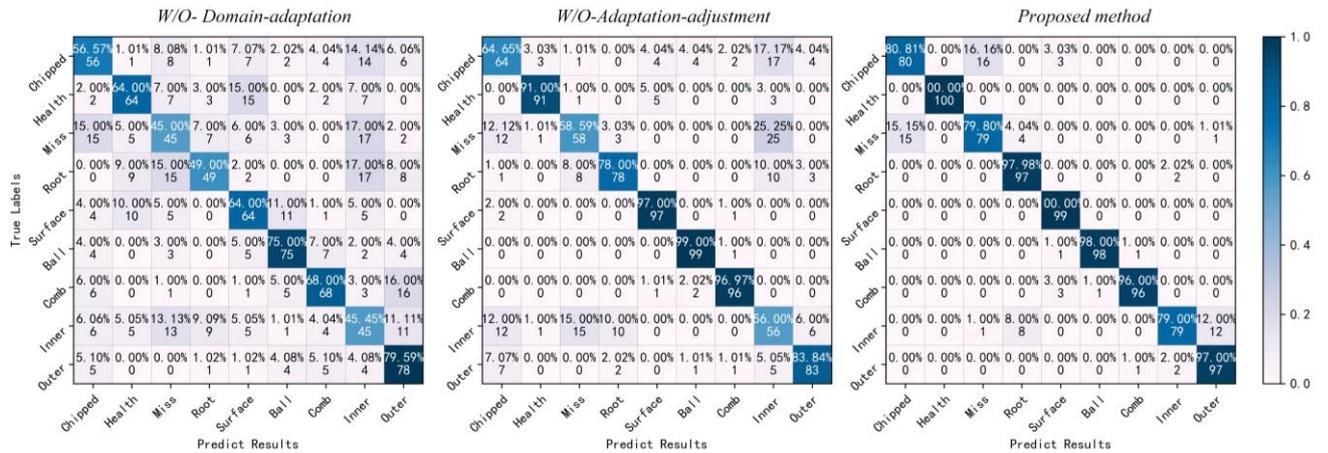

Fig. 7 Confusion matrix of test results for group C.

## 4. CONCLUSION

Fault diagnosis in mechanical equipment plays a pivotal role in supporting industrial production. Notably, variations in speed, load, and other factors during mechanical equipment operation lead to significant differences in data distribution, posing challenges for fault diagnosis. Moreover, conventional cloud-based fault diagnosis methods often encounter issues such as time delays and data security concerns during application deployment, while standard fault diagnosis methods for cross-operating conditions cannot be directly applied to edge computing devices. Therefore, fault diagnosis under diverse operating conditions using edge computing holds significant research value. This study proposes a lightweight fault diagnosis framework based on domain adaptation for edge computing scenarios. The objective is to achieve accurate fault diagnosis under cross-working conditions while maintaining real-time diagnosis capabilities. By incorporating domain adaptation learning into fault diagnosis, it aligns feature distributions across different domains in a high-dimensional feature space, thereby identifying common feature spaces among diverse domains. Leveraging a knowledge transfer approach, fault diagnosis expertise obtained from the cloud-based C-model is transferred to the lightweight E-model. We conducted validation experiments on two devices. The proposed method significantly enhances diagnostic accuracy, with average improvements of 34.44% compared to existing methods, respectively. Additionally, our method achieves an average increase in inference speed of 80.47% compared to contrasting methods. Compared to the C model, the parameter count of the E model is reduced by 96.37%, and Flops are reduced by 83.08%. Furthermore, we conducted ablation experiments, and the diagnostic accuracy of the proposed method was improved by an average of 22.67% and 11.99% compared to the method lacking domain adaptation and the method lacking adaptation, respectively.

In summary, the domain adaptation method effectively improves the effectiveness of knowledge transfer, enabling the transfer of fault diagnosis knowledge learned under a single operating condition to actual operational conditions. This provides an effective approach for cross-condition fault diagnosis in edge computing scenarios, further promoting the widespread application of intelligent fault diagnosis in practical scenarios. However, although the proposed method achieves excellent lightweight effects, it still requires fine-tuning on the edge side with a small amount of operational data during knowledge transfer, which increases the overhead at the edge to some extent. Future work could explore the design of an online learning mechanism, allowing the model to be updated and adjusted based on real-time data collected on edge devices and then deployed[37]. Through online learning, the model can continuously adapt to changes in the edge environment, achieving adaptive capabilities without the need for fine-tuning.

## Declaration of competing interest

The authors declare that they have no known competing financial interests or personal relationships that could have appeared to influence the work reported in this paper.

## Funding

This research was supported in part by the National Key R&D Program of China Young Scientists Project [Grant number 2022YFC2204700].

## Data availability

Data will be made available on request.